\documentclass{article}
\pdfoutput=1
\usepackage[margin=3cm]{geometry}
\usepackage{comment}
\usepackage{amsmath,amssymb,extarrows,mathtools,graphicx,subfigure,setspace,bbold}
\usepackage{cite}
\usepackage{mcite}
\usepackage{slashed}
\usepackage{color}
\usepackage{xargs}                      
\usepackage{tikz}

\usetikzlibrary{decorations.pathreplacing}
\makeatletter
\def\vhrulefill#1{\leavevmode\leaders\hrule\@height#1\hfill \kern\z@}
\makeatother
\makeatother
\renewcommand{\baselinestretch}{1}
\newcommand{\beq}{\begin{equation}}
\newcommand{\eeq}{\end{equation}}
\newcommand{\be}{\begin{equation}}
\newcommand{\bea}{\begin{eqnarray}}
\newcommand{\eea}{\end{eqnarray}}
\newcommand{\ba}{\begin{array}}
\newcommand{\ea}{\end{array}}
\newcommand{\ee}{\end{equation}}
\newcommand{\bes}{\begin{equation*}}
\newcommand{\beas}{\begin{eqnarray*}}
\newcommand{\eeas}{\end{eqnarray*}}
\newcommand{\bas}{\begin{array*}}
\newcommand{\eas}{\end{array*}}
\newcommand{\ees}{\end{equation*}}
\newcommand{\nn}{\nonumber}
\newcommand{\pd}{\partial}

\def\K{{\mathcal K}}
\def\bg{{\bar g}}
\def\bR{{\bar R}}
\def\bgamma{{\bar{\gamma}}}
\def\a{\alpha}
\def\b{\beta}

\def\ve{\varepsilon}

\def\l{\lambda}

\def\m{\mu}
\def\n{\nu}

\def\r{\rho}
\def\s{\sigma}

\def\fr{\frac}

\def\pa{\partial}

\def\nn{\nonumber}

\def\lt{\left}
\def\rt{\right}
\def\({\left(}
\def\){\right)}
\def\[{\left[}
\def\]{\right]}

\numberwithin{equation}{section}

\begin{document}
\begin{titlepage}
\begin{flushright}\vspace{-3cm}
{
IPM/P-2016/012 \\
\today }\end{flushright}
\bigskip\bigskip
\begin{center}
{\Large{\bf{Holographic Entanglement Entropy, Field Redefinition Invariance 
and Higher Derivative Gravity Theories}}}

\bigskip\bigskip\bigskip

\centerline{\large{\bf{M.R. Mohammadi Mozaffar, A. Mollabashi, M.M. Sheikh-Jabbari, M.H. Vahidinia\footnote{e-mails:  m\_mohammadi, mollabashi, vahidinia@ipm.ir, 
jabbari@theory.ipm.ac.ir}}
}}

\normalsize
\bigskip\medskip
{\it School of Physics, Institute for Research in Fundamental
Sciences (IPM),\\ P.O.Box 19395-5531, Tehran, Iran}

\end{center}
\setcounter{footnote}{0}

\begin{abstract}
\noindent

It is established  that physical observables in local quantum field theories should be invariant under invertible field redefinitions. It is then expected that this statement should be true for the entanglement entropy and moreover that, via the gauge/gravity correspondence, the recipe for computing entanglement entropy holographically should also be invariant under local field redefinitions in the gravity side. We use this fact to fix the recipe for computing holographic entanglement entropy (HEE) for $f(R,R_{\mu\nu})$ theories which could be mapped to Einstein gravity. An outcome of our prescription is that the surfaces that minimize the corresponding HEE functional for $f(R,R_{\mu\nu})$ theories always have vanishing trace of extrinsic curvature and that the HEE  may be evaluated using the Wald entropy functional. We show that similar results follow from the FPS and Dong HEE functionals, for Einstein manifold backgrounds in $f(R,R_{\mu\nu})$ theories.

\end{abstract}

\end{titlepage}
\renewcommand{\baselinestretch}{1.1}  
\tableofcontents

\section{Introduction}

Hilbert space of local field theories is specified with all possible field configurations on a given constant time slice. Alternatively, via the operator-state correspondence, one may use the set of all possible local operators at a given neighbourhood of compact support to specify the Hilbert space. One of the implications of the previous statement, which becomes more clear in the path integral formulation of quantum field theories, is \emph{field redefinition invariance} of physical observables. Explicitly, let us consider a local field theory described with generic set of fields $\Phi(x)$. We may  choose to describe the same theory with $\tilde\Phi(x)=\tilde\Phi[\Phi(x)]$. Here we will always assume that this field redefinition map is invertible, i.e. the Jacobian of the transformation is nonzero.  The field redefinition invariance then states that  there should be a one-to-one correspondence between physical observables of 
the theory in $\Phi$ and in $\tilde\Phi$-descriptions; field redefinitions are to be viewed as change of basis in the Hilbert space and the physical observables should be independent of the choice of basis. A specific but very important case of the field redefinition invariance concerns the S-matrix elements which are related to the set of $n$-point $(n\geq 2)$ functions  of all local operators through the standard LSZ theorem, see \cite{Weinberg:1995mt} for the proof and further discussions. 

Local quantum field theories have local or non-local physical observables. The best known non-local physical observables are Wilson or Polyakov loops in gauge theories. However, recent developments have emphasized on entanglement or Renyi entropy or other information theoretic quantities, as other non-local physical observables Entanglement entropy is an observable\footnote{Of course entanglement and Renyi entropies are state-dependent observables exactly in the same sense that one calls expectation values as observables in a quantum theory.} (see for example \cite{Nishioka:2009un}). Since in this work we will be mainly concerned with the entanglement entropy let us focus on this non-local observable. The entanglement entropy is computed between two parts of the Hilbert space. For example, let us suppose that the Hilbert space ${\cal H}$ is divided into two mutually exclusive parts ${\cal H}_A$ and ${\cal H}_B$, i.e. $\mathcal{H}=\mathcal{H}_A\otimes\mathcal{H}_B$. One may then define a reduced density matrix over ${\cal H}_A$, $\rho_A$, by integrating out states in ${\cal H}_B$.  The entanglement entropy between $A$ and $B$ parts is then defined as the von Neumann entropy of $\rho_A$. For a given local field theory, whose Hilbert space is specified by field configurations on constant time slices, there is a one-to-one relation between points on the constant time slice and the Hilbert space. In this case ${\cal H}_A$ and ${\cal H}_B$ regions can have a direct geometric realization: Let us divide  the constant time slice into two regions $V_A$ and $V_B$ which are separated by a codimension two surface $\partial V_A$. A well known and established result in the local free quantum field theories is that the leading term in the entanglement entropy between these two regions is proportional to the area of $\partial V_A$, ${\cal A}$, in units of the UV cutoff of the theory \cite{Bombelli:1986rw, *Srednicki:1993im}. 
This result which is obtained for free field theory with no excitations can be extended to weakly coupled theories and with generic perturbative excitations on the background field theory. Away from the weakly coupled fixed point, there is no direct  field theory computation (applicable to generic field theories, beyond 2d CFT's) for calculating the entanglement entropy. 

Following the steps of computation of entanglement entropy for local quantum field theories, one can check that like the local observables, entanglement entropy, mutual and multipartite information, and relative entropy should also be invariant under local field redefinitions, as field redefinition is a change of basis in the Hilbert space and as long as it does not tamper with the geometric partitioning of the Hilbert space through dividing a constant time slice into two regions, it should not alter the outcome of trace over $B$ and then $A$ parts.

On the other hand, one may use the AdS/CFT setting (or its ``more relaxed version'' gauge/gravity correspondence) to compute physical observables in strongly coupled field theories (at large $N$). Ryu and Takayanagi (RT) in their seminal paper \cite{Ryu:2006bv} made the first such proposal for computing entanglement entropy holographically. The RT proposal was extended to non-static  backgrounds (the HRT proposal) \cite{Hubeny:2007xt}. The essence of RT (or HRT) proposal is very elegant and simple: The entanglement entropy associated with a region $V_A$ at the AdS boundary is the area of the minimal (or extremal) surface whose boundary is anchored to the AdS boundary at $\partial V_A$, divided by $4G_N$. This proposal can be extended to any asymptotically AdS geometry, in particular asymptotically AdS black holes, corresponding to field theories at non-zero temperature \cite{Ryu:2006bv, Ryu:2006ef}. The (H)RT proposal has successfully passed several non-trivial checks \cite{Ryu:2006ef, Headrick:2007km, Hubeny:2007xt} and has been convincingly argued to be the right proposal for computing entanglement entropy within the AdS/CFT framework \cite{Lewkowycz:2013nqa}.

The RT and HRT proposals are made for Einstein gravity theory (plus minimally coupled matter fields). However, it is a well-known fact that in a semi-classical regime the Lagrangian of gravity theories would involve higher derivative terms, higher powers of Riemann curvature (and in general the derivatives of Riemann).  These corrections within string theory setting, after using (super)gravity field redefinitions,  can generically be brought to the form of  Lovelock theories  order by order in an $\alpha'$ expansion \cite{Zwiebach:1985uq, *Zumino:1985dp,*Jack:1987pc}. In spite of the fact that higher derivative gravity theories generically have ghosts, it is of theoretical interest to extend the Holographic Entanglement Entropy (HEE) recipes to generic higher derivative cases. There have been several papers already on the topic and there are some different proposal and recipes, e.g. the FPS proposal \cite{Fursaev:2013fta},  de Boer et al. and Hung et al. proposal \cite{deBoer:2011wk, Hung:2011xb}, the proposal by Dong and Camps \cite{Dong:2013qoa, Camps:2013zua} and other works on the topic e.g. \cite{Myers:2013lva,*Bhattacharyya:2013jma,*Pourhasan:2014fba, *Alishahiha:2013dca, Chen:2013rcq, Miao:2013nfa, Bhattacharyya:2013gra, Bhattacharyya:2014yga, Miao:2014nxa, Camps:2014voa}. 

One of the fundamental statements in the AdS/CFT duality is that the VEV of operators in the field theory is related to the value of the bulk (gravity) fields at the boundary of AdS \cite{Witten:1998qj} (see also the first section of the lecture notes \cite{Sheikh-Jabbari:2013xx}). This picture then implies that field redefinition in the field theory side, which is the same as a change of basis in the space of field theory operators,  should translate into a field redefinition in the gravity side. Of course, while the precise mapping between the two field redefinitions could be complicated, for what concerns us here, the existence of a field redefinition in the gravity side is enough. Recalling the above statements in the field theory side one would then expect that the RT or HRT proposals, which are recipes for computing entanglement entropy in the gravity side,  should also be field redefinition invariant. 

Under a generic field redefinition in a gravity theory the form of the action would in general change. For example, it is well known that a generic $f(R)$ theory of gravity can be mapped onto a scalar-Einstein theory upon an appropriately chosen field redefinition \cite{Magnano:1993bd}. Of course, this mapping and field redefinition is well-defined if the map is invertible and one should check this condition on each solution of the theory to ensure that we have a well-defined mapping. One may then use the above line of reasoning to extend the basic RT and HRT proposals which were made for Einstein gravity theory, to more general gravity theories. Explicitly our guiding principle is that: For any higher derivative gravity theory which can be mapped to Einstein gravity upon a field redefinition we have a way  to fix the HEE recipe. Here we will use the terminology of ``Einstein-like'' theory, for the gravity theories which could be mapped to Einstein gravity upon field redefinitions.  In other wording, our recipe is to map an ``Einstein-like'' theory to the Einstein frame and use RT or HRT proposals and map it back to the original theory. In this way we get the HEE proposal for such higher derivative gravity theories, using the RT or HRT proposals as reference points. This procedure in depicted in the above figure.

Before moving to the specific application and exploration of the above guiding principle, it is worth noting a parallel and closely related discussion on the black hole (thermal) entropy. In his seminal paper \cite{Wald:1993nt}, R. Wald gave a general formula for computing the entropy associated with a black hole solution to a generic gravity theory the action of which is a general diffeomorphism invariant functional of Riemann curvature and its covariant derivatives,  and the metric. In this formulation black hole entropy was introduced as the Noether-Wald charge associated with the Killing vector field generating the (Killing) horizon. In a paper which was published shortly after Wald's paper \cite{Jacobson:1993vj}, it was argued that black hole and hence Wald's entropy formula, should be invariant under field redefinitions of the sort discussed above (see also \cite{Koga:1998un}). So, in this sense our new input here is to extend similar idea to more general case of the (holographic) entanglement entropy. We note that, horizon is in fact a particular minimal (extremal) surface with vanishing  extrinsic curvature; the horizon can be a disconnected part of a minimal (extremal) surface \cite{Hubeny:2013gta,*Azeyanagi:2007bj}. This latter, among other things, provides another cross-check for any proposal for computing HEE in generic higher derivative gravity theory: the ``HEE functional'' (as it is known in the terminology of the field) should reduce to Wald's entropy formula when computed on the Killing horizons. 

The organization of this paper is as follows. In section  \ref{sec-2}, we consider general modified gravity theories with Lagrangians of the form $f(R,R_{\mu\nu})$ and discuss the field redefinition which maps the theory onto Einstein-Hilbert theory. In section \ref{sec-3}, we use the field redefinitions of the previous section to fix the HEE functional for $f(R,R_{\mu\nu})$ theories. In section \ref{sec-4}, we compare our proposal and the other proposals for computing HEE in the literature.  The last section is devoted to open questions and the outlook. In two Appendices we have provided some more detailed analysis on the comparison between our method and the functionals existing in the literature and also discussed some  examples to show how our method works explicitly and to show the matching with the other recipes.

\begin{figure}[t]\label{fig1}
\begin{center}
\includegraphics[scale=1]{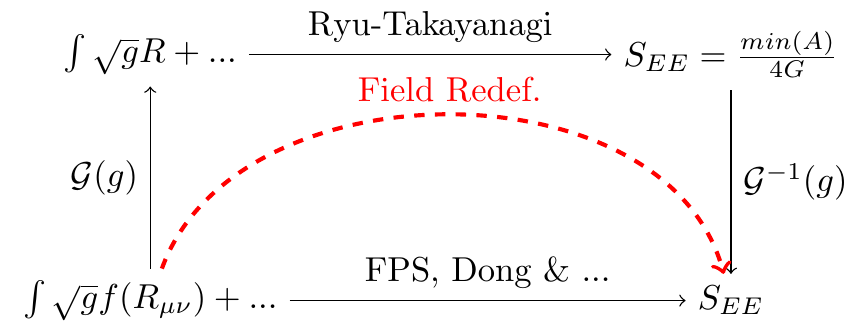}
\end{center}
\end{figure}

\section{$f(\text{Ric})$ theories are ``Einstien-like'', a review}\label{sec-2}

In this work we will consider metric-compatible, torsion-free connections, within metric formulation of gravity. We will briefly discuss more general cases in the discussion section. In this case,  Riemann curvature has $d^2(d^2-1)/12$ independent components in $d$ dimensions and the Riemann curvature may be decomposed into Ricci curvature $R_{\mu\nu}$ and the Weyl curvature $W_{\alpha\beta\mu\nu}$. Ricci tensor is symmetric and has $d(d+1)/2$ components and the rest of $d(d+1)(d+2)(d-3)/12$ components are in Weyl tensor, which is traceless over all indices. 

In this section we explore a family of higher derivative gravity theories which could be mapped to Einstein gravity upon appropriate field redefinitions, i.e. the higher derivative theories which fit within our definition of ``Einstein-like'' theories. The above simple counting, noting that metric is also a symmetric rank two tensor and has the same number of degrees of freedom as Ricci tensor, suggests that one should be able to  map $f(\text{Ric})$ onto Einstein-Hilbert Lagrangian.  This area has of course been explored in many papers in the literature, e.g. in  \cite{Magnano:1993bd, Sotiriou:2008rp, Baykal:2013gfa, *Rodrigues:2011zi,*Chiba:2005nz}. 
In this section we review such mappings, as discussed in the aforementioned papers.

\subsection{From $f(R)$ to Einstien-Hilbert}\label{sec-2.1}
The most general action for a $f(R)$ gravity is given by
\bea\label{action1}
I=\frac{1}{2\kappa^2}\int d^{d+1}x \sqrt{|g|}\left[f(R)+ {\cal L}_{matter}(j^1g_{\mu\nu}, \phi)\right],
\eea
where $\phi$ in the second term denotes generic matter fields and the matter sector do not contain second derivative of the metric.\footnote{$j^k T$ means $k$th order jet prolongation of $T$. Roughly, it denotes the set of $T$ and partial derivatives of $T$ up to $k$th order.} The equation of motion for the above action is given by
\bea\label{eomfofR}
\left(R_{\mu\nu}+g_{\mu\nu}\Box-\nabla_\mu\nabla_\nu\right)f'(R)-\frac{1}{2}g_{\mu\nu}f(R)=T_{\mu \nu},
\eea
where $T_{\mu \nu}=-\frac{1}{\sqrt{g}}\frac{\delta \sqrt{g}\mathcal{L}_{matter}}{\delta g^{\mu \nu}}$. We then introduce a new scalar field $\Phi$, such that $\Phi=R$. This latter may be added to the action using a Lagrange multiplier $\lambda$, leading to action in which the equation of motion for $\Phi$ is given by $\lambda=f'(\Phi)$. Next, after the Weyl scaling,
\bea\label{scalingmetric}
g_{\mu \nu}=e^{2\sigma} \bg_{\mu \nu},
\eea
the action \eqref{action1} becomes
$$ 
\hspace*{-5mm}I=\frac{1}{2\kappa^2}\int d^{d+1}x\;e^{(d+1)\sigma}\sqrt{|\bg|}\left[f(\Phi)-\Phi f'(\Phi)+f'(\Phi)e^{-2\sigma}\left(\bR-2d\nabla^2\sigma-d(d-1)\partial_\mu \sigma \partial^\mu \sigma\right)+{\cal L}_{matter}(j^1\bar g_{\mu\nu}, \sigma, \phi)\right],
$$
where covariant derivative $\nabla$ is defined for metric $\bg_{\mu\nu}$.
In order to extract the Einstein part of the above action one may consider the following redefinitions
\bea\label{f}
f'(\Phi)=\frac{\kappa^2}{8 \pi G_{d+1}}e^{-\sqrt{\frac{d-1}{2d}}\varphi},\qquad \varphi=\sqrt{2d(d-1)}\sigma\,,
\eea
which leads to
\bea\label{action4}
I&=&{\frac{1}{16 \pi G_{d+1}}}\int d^{d+1}x\;\sqrt{|\bg|}\left[\bR-\frac12\partial_\mu \varphi\partial^\mu \varphi- V(\varphi)+{\hat{\cal L}}_{matter}(\bg,\partial\bg, \varphi; \phi)\right],\cr
V(\varphi)&=&{\frac{8 \pi G_{d+1}}{\kappa^2}} e^{\sqrt{\frac{(d+1)^2}{2d(d-1)}}\varphi}\left(\Phi f'(\Phi)-f(\Phi)\right), 
\eea
where from \eqref{f} one may solve $\Phi$ for $\varphi$. Some brief comments are in order \cite{Sotiriou:2008rp}:
\begin{itemize}
\item  $\kappa$ is not necessarily directly related to $d+1$ dimensional Newton constant, $G_N$ by definition is the coefficient in front of the Einstein-Hilbert gravity action. In particular, we note that $G_N$ is related to $\kappa$ through \eqref{f} and that it is positive if $f'(\Phi)$ is positive on a given profile for $\Phi$. This latter is of course determined by the field equations; acceptable solutions are those with $f'(\Phi)>0$.

\item The equivalence of $f(R)$ and Einstein-scalar theory is an on-shell statement, i.e., it maps any given solution of the $f(R)$ theory \eqref{action1} to Einstein-scalar theory \eqref{action4}. Importantly, one should check that the map is invertible (on the solutions of the e.o.m.).

\item For the special case of $f(\Phi)=\lambda\Phi^n$, 
$$
V(\varphi)=V_0 e^{\beta\varPhi},\qquad \beta=\frac{2n-d-1}{(n-1)\sqrt{2d(d-1)}},\ \ V_0=\frac{n-1}{n}\left(\frac{\kappa^2}{8 \pi n G_{d+1} \lambda}\right)^{\frac{1}{n-1}}.
$$
For the special case of Einstein-Hilbert action with $n=1$, as we expect, the potential is simply zero and the scalar $\varPhi$ should also be set to a constant (\emph{cf.} \eqref{f}). For $f(\Phi)=\lambda\Phi^n, n\geq 1$ cases, background solutions with $R=0$ would correspond to $\varphi=\infty$ cases and the map in not invertible; i.e. these theories may have solutions which are not captured through our mapping. The $R^2$ case has been recently studied in \cite{Alvarez-Gaume:2015rwa}.

\end{itemize}

\subsection{From $f(\text{Ric})$ to Einstein-Hilbert}

Consider the following  gravity action
\beq \label{fAction}
I=\frac{1}{2\kappa^2}\int d^{d+1}x \sqrt{|g|} \left[f(\text{Ric})+{\cal L}_{matter}(j^1 g_{\mu \nu},\Phi)\right]
\eeq
where $f(\text{Ric})\equiv f(R_{\mu \nu},g_{\mu \nu})$ is the most general scalar function which only depends on Ricci tensor and metric. Therefore, $f(\text{Ric})=f(R^\mu_{\ \nu})$ and $\frac{\partial_{f}}{\partial{g_{\mu\nu}}}$ and $\frac{\partial_{f}}{\partial{R_{\mu\nu}}}$ are related to each other by a factor of Ricci tensor. We also consider the matter fields contribution ${\cal L}_{matter}$ due to some matter fields $\Phi$ (not necessary scalars). In general, this action leads to a fourth order field equation for metric $g_{\mu \nu}$ 
\beq \label{fEq}
 \frac{\pd f }{\pd g^{\mu \nu}}-\frac{1}{2} f g_{\mu \nu}+\frac{1}{2}\left(g_{\mu \nu} \nabla_\lambda \nabla_\rho-g_{\mu \rho} \nabla_\nu \nabla_\lambda -g_{\nu \rho}  \nabla_\lambda \nabla_\mu+g_{\lambda \nu}g_{\rho \mu}\Box\right)\frac{\pd f}{\pd R_{\lambda \rho}}=T_{\mu \nu},
\eeq
where $T_{\mu \nu}=-\frac{1}{\sqrt{g}}\frac{\delta \sqrt{g}\mathcal{L}_{matter}}{\delta g^{\mu \nu}}$ and the covariant derivative $\nabla$ is compatible with the metric $g_{\mu \nu}$.
It is well-known that \cite{Magnano:1987zz} one can redefine metric field using 
\begin{eqnarray} \label{gredef}
\bar{g}^{\mu \nu}&=&\left(|\text{det} \frac{ \pd \sqrt{|g|}f }{\pd R_{\mu \nu}} | \right)^{\frac{-1}{d-1}} \frac{\pd \sqrt{|g|}f }{\pd R_{\mu \nu}} \nonumber\\
&=& |g|^{\frac{-1}{d-1}}| \text{det} ({\cal F}^{\mu\nu})|^{\frac{-1}{d-1}} {\cal F}^{\mu\nu},
\end{eqnarray}
where
\beq\label{F-g-f(Ric)}
{\cal F}^{\mu\nu}\equiv \frac{\pd f}{\pd R_{\mu \nu}}.
\eeq
To obtain a new gravity action
\beq \label{newAction}
I={\frac{1}{16\pi G_{d+1}}}\int d^{d+1}x \sqrt{|\bg|} \left[\bar{R} +{\cal L}_{matter}(j^1 \bar{g},j^1 g,\Phi)\right]
\eeq
corresponding to the Einstein theory and in the presence of extra matter field associated to the original metric $g_{\mu \nu}$. This action leads to a second order field equation for the new metric $\bar{g}_{\mu \nu}$ \cite{Magnano:1987zz}.
Note that the ${\cal L}_{matter}(j^1 \bar{g},j^1 g,\Phi)$ term now contains terms like 
\begin{align}
\begin{split}
\bar{g}^{\a\b}\left({\mathcal{Q}^\r}_{\s\a}{\mathcal{Q}^\s}_{\r\b}-{\mathcal{Q}^\r}_{\a\b}{\mathcal{Q}^\s}_{\s\r}\right),
\qquad {\mathcal{Q}^\a}_{\b\n}=\frac{g^{\a\s}}{2}\left(\bar{\nabla}_\n g_{\s\b}+\bar{\nabla}_\b g_{\s\n}-\bar{\nabla}_\s g_{\a\b}\right).
\end{split}\nonumber
\end{align} 
Therefore, the action \eqref{newAction} describes two spin two fields, a massless graviton represented through $\bar g_{\mu\nu}$ and a generically massive spin two coming from $g_{\a\b}$. This latter may generically have negative kinetic term (a ghost-like degree of freedom).

Given the above mapping some comments are in order:
\begin{itemize}
\item For the Einstein-Hilbert action, one may readily check that, as expected, \eqref{gredef} reduces to $\bar g_{\mu\nu}=g_{\mu\nu}$.
\item One can show that in the case of $f(\text{Ric})=f(R)$ curvature tensors  the above procedure reduces to what was discussed in Sec. \ref{sec-2.1}. In this case one finds ${\cal F}^{\mu\nu}=f'(R)g^{\mu\nu} $ and by using \eqref{gredef} we have $\bar{g}_{\mu \nu}=f'(R)^{\frac{2}{d-1}}{g}_{\mu \nu}$. On the other hand combining \eqref{scalingmetric} and \eqref{f} and replacing $\kappa^2$ with $8\pi G_{d+1}$ in \eqref{newAction} leads to
$\bar{g}_{\mu \nu}=f'(R)^{\frac{2}{d-1}}{g}_{\mu \nu}$.

\item Note that $\bar g_{\mu\nu}$ is governed by a second order field equation (the usual Einstein equation). In fact, through introducing the new field  $\bar g_{\mu\nu}$, we have doubled dynamical fields to $\bar g_{\mu\nu}$  and $g_{\mu\nu}$ and the fourth order field equation \eqref{fEq} is now replaced with two second order equations.

\item Recalling the discussion in the opening of this section, we learn that we cannot replace Weyl or Riemann curvature tensors by a field redefinition of the kind used for $f(R)$ or $f$(Ric) theories, as Riemann and Weyl tensors are not in 
simple spin $s$ representation of the Lorentz group.

\end{itemize}

\section{Field redefintions and holographic entanglement entropy}\label{sec-3}

\paragraph{The RT proposal:} For a $d$ dimensional CFT with an \emph{Einstein-Hilbert gravity} dual, the entanglement entropy $S_E$ associated with a region $V_A$ bounded with the compact $d-2$ dimensional space $\partial V_A$, may be computed ``holographically''  minimizing the functional \cite{Ryu:2006bv}
\beq\label{RT-functional}
S_{EE}=\frac{1}{4G_{d+1}} \int_{\gamma_{A}}d^{d-1}\xi \sqrt{\bar{\gamma}},
\eeq
over the $d-1$ dimensional spacelike surfaces $\bar{\gamma}$ which intersect causal boundary of AdS$_{d+1}$ on $\partial V_A$. $\bar{\gamma}_{ab}$  denotes the induced metric on the minimal surface $\bar{\gamma}$. The minimal surface $\bar{\gamma}$ may be denoted through the embedding equations 
$$
x^\mu=x^\mu(\xi^a), \qquad \mu=1,2,\cdots d,\ a=1,2,\cdots, d-1
$$
or alternatively through $\Psi^i(x^\mu)=0$ which denotes a $d-1$ dimensional surface with normal vector $n^i_\mu=\partial_\mu\Psi^i$. The induced metric is then
\beq \label{indmet2}
\bar{\gamma}_{ab}=g_{\mu \nu}\frac{\partial x^{\mu}}{\partial \xi^{a}}\frac{\partial x^{\nu}}{\partial \xi ^{b}},
\eeq
or alternatively is given through $g_{\mu\nu}-n^i_\mu n^i_\nu$. The RT proposal prescribes minimizing the functional \eqref{RT-functional} to fix $\bar{\gamma}$. This latter states that the RT minimal surface should have vanishing  trace of extrinsic curvature.

The RT proposal for the Holographic Entanglement Entropy (HEE) may be used for any static Asymptotically AdS$_{d+1}$ (AAdS$_{d+1}$) backgrounds and the minimal surfaces $\bar{\gamma}$ reside on constant time slices in these backgrounds. The shape of the minimal surface, the induced metric $\bar{\gamma}_{ab}$, and hence the entanglement entropy $S_E$ depend on the $\partial V_A$ as well as the background metric.  This proposal has been used to computed the entanglement entropy for various entangling region shapes (e.g. a sphere, strip or cylinder) and many AAdS backgrounds e.g. AdS-black holes \cite{Ryu:2006ef}. Moreover, the RT proposal has been extended to the ``HRT proposal'' for non-static and stationary backgrounds \cite{Hubeny:2007xt}. In the latter cases the minimal surface is replaced with an ``extremal surface'' with a very similar construction.

We stress again that this functional only works for Einstein theory of gravity with possibly other minimally coupled matter fields. In the rest of this section, we show that one may use field redefinition invariance to obtain a proposal for computing the HEE for more general Einstein-like theories. The general framework is of course remaining the same as the RT, namely the HEE is obtained through minimizing a functional  over a minimal surface, and our task is to find the functional. In what follows by ``minimal surface" we refer to the surface which minimizes the corresponding functional  for HEE (not necessarily the area functional).

\subsection{Holographic entanglement entropy functional for $f(R)$} \label{sec-3.1}

Let us suppose that there exists a $d$ dimensional (conformal) field theory which is dual to an $f(R)$ gravity theory coupled to some matter fields, with the action of the form \eqref{action1}.  
To obtain the functional for computing HEE for this theory we note that we may use the RT proposal for the equivalent Einstein-scalar theory \eqref{action4}. This will then allow us to provide the HEE functional directly for the $f(R)$ theory. Explicitly, consider the RT functional for theory \eqref{action4} as the starting point
\bea
S_{EE}=\frac{1}{4G_{d+1}}\int d^{d-1}\xi \sqrt{\bar \gamma}.
\eea
Now using \eqref{scalingmetric} and \eqref{indmet2}, we can replace induced metric $\bgamma$ with $\gamma$ where the latter one is defined on background $g_{\mu \nu}$. Using this fact, the holographic entanglement entropy for $f(R)$ theory is given by
\begin{eqnarray}\label{func-fofr}
S_{EE}&=&\frac{1}{4G_{d+1}}\int d^{d-1}\xi\;e^{(1-d)\sigma} \sqrt{\gamma} \nn \\
&=&\frac{4\pi}{\kappa^2}\int d^{d-1}\xi\;f'(R) \sqrt{\gamma},
\end{eqnarray}
where in the last equality we used \eqref{f}. As we see the function $f$ appears directly in the HEE functional and hence enters directly in the minimal surface $\gamma$. This result was previously noted in \cite{Dong:2013qoa}.

One may alternatively minimize the functional \eqref{func-fofr} to find the minimal surface and by evaluating the function over this minimal surface compute the HEE. One can then readily see that this minimal surface should have vanishing trace of extrinsic curvature for constant curvature solutions.

\subsection{Holographic entanglement entropy functional for $f(\text{Ric})$} \label{sec-3.2}
We can generalize the previous analysis to obtain entanglement entropy functional for an $f(\text{Ric})$ theory from the RT functional. Starting from \eqref{gredef}, we learn that
\begin{eqnarray} \label{gamma-bar-g}
{\bgamma}={{|g|} \det{({\cal F}^{\mu\nu}})}\ {\det{({\cal F}^{-1}_{ab})} },
\end{eqnarray}
where ${\cal F}^{-1}_{ab}$ is the induced-inverted ${\cal F}^{\mu\nu}$, explicitly
\beq\label{F-inv-induced}
{\cal F}^{-1}_{ab}\equiv ({\cal F}^{-1})_{\mu\nu} \frac{\partial x^\mu}{\partial\xi^a}\frac{\partial x^\nu}{\partial\xi^b}, \qquad {\cal F}^{-1}_{\mu\nu} {\cal F}^{\nu\rho}=\delta_{\mu\ }^{\rho}.
\eeq
Note in the above, the induced metric $\bgamma$ should be viewed as a function of the transformed (redefined) metric $\bg$ defined in \eqref{gredef} and that the latter is a function of ${\cal F}^{\mu\nu}$ \eqref{F-g-f(Ric)}. Using \ref{gamma-bar-g}, one finds the HEE functional in $f(Ric)$ theory as
\beq
S_{EE}=\frac{1}{4G_{d+1}}\int_{\gamma_{A}}\sqrt{{{|g|} \det{({\cal F}^{\mu\nu}})}\ {\det{({\cal F}^{-1}_{ab})} }}
\eeq

Although we implicitly have our proposed functional, one cannot proceed further analytically for general cases. We shall hence restrict ourselves to two cases: 
\begin{enumerate}
\item Cases which are perturbations around Einstein-Hilbert theory, with $f(\text{Ric})=R-2 \Lambda +\lambda \tilde{f}(\text{Ric})+ {\cal O}(\lambda^2)$. 
\item The general $f(\text{Ric})$ theory which admits an Einstein manifold solution and compute the HEE for these backgrounds for which $R_{\mu \nu}=\frac{R}{d+1}g_{\mu \nu}$.
\end{enumerate}

\paragraph{1. Perturbative $f(\text{Ric})$.}
Let us assume $f(\text{Ric})$ as a modified version of Einstein-Hilbert action where higher curvature terms appear as a perturbation and keep this correction up to the first order
\beq
f(R_{\mu \nu},g_{\mu \nu})\equiv R-2\Lambda+\lambda \tilde{f} (R_{\mu \nu},g_{\mu \nu})+\mathcal{O}(\lambda^2).
\eeq
For this case
\begin{eqnarray}\label{gredef2}
{\cal F}^{\mu\nu}= g^{\mu \nu} +\lambda \frac{\pd\tilde{f}}{\pd R_{\mu \nu}},\qquad 
\text{det} {\cal F}^{\mu\nu}=\mid g\mid^{-1} \left(1+\lambda \tilde{{\cal F}} \right)+\mathcal{O}(\lambda^2),
\end{eqnarray}
where $\tilde{{\cal F}}= g_{\mu \nu}\frac{\pd\tilde{f}}{\pd R_{\mu \nu}}$. Using this one may obtain
\beq
\bar{g}_{\mu \nu}= {g}_{\mu \nu}-\lambda \left(\tilde{{\cal F}}_{\mu \nu}-\frac{1}{d-1} {g}_{\mu \nu} \tilde{{\cal F}}\right),\qquad \tilde{{\cal F}}_{\mu\nu}\equiv g_{\mu\alpha}g_{\nu\beta}\frac{\pd\tilde{f}}{\pd R_{\alpha \beta}}.
\eeq
Using \eqref{gamma-bar-g} and the above, one may calculate $\sqrt{\bar{\gamma}}$:
$$
\sqrt{\bar{\gamma}}=\sqrt{\gamma}\big(1+\frac{\lambda}{2}(\tilde{{\cal F}}-\tilde{{\cal F}}_{ab}\gamma^{ab})\big)
$$
where $\tilde{{\cal F}}_{ab}$ is the tensor $\tilde{{\cal F}}_{\mu\nu}$ induced on the minimal surface and $\gamma^{ab}$ is the inverse of induced metric $\gamma_{ab}$. Since the theory with $\bg$ metric is Einstein-Hilbert, one may use the standard RT functional, yielding
\begin{subequations}\label{pertFunctional}\begin{align}
S_{EE}&=
\frac{1}{4G_{d+1}} \int_{\gamma_{A}}d^{d-1}\xi\sqrt{\gamma}\left[1+ \frac{\lambda}{2} \big(\tilde{{\cal F}}-\tilde{{\cal F}}_{ab}\gamma^{ab}\big)\right]\nonumber\\
&= \frac{1}{4G_{d+1}} \int_{\gamma_{A}}d^{d-1}\xi\sqrt{\gamma}\left[1+ \frac{\lambda}{2} \tilde{{\cal F}}^{\mu\nu} n^{(i)}_\mu n^{(i)}_\nu\right]
\end{align}\end{subequations}
where $n^{(i)}_\mu$ is the normalized vector normal to the minimal surface.

\paragraph{2. Einstein manifold solutions $f(\text{Ric})$.} We are usually interested in computing the HEE on the background metric solutions which are Einstein manifolds, i.e.
\beq\label{Einstein-manifold}
R_{\mu \nu} = \frac{R}{d+1}g_{\mu \nu},
\eeq
where $R$ is a constant Ricci scalar. Using the Bianchi identity one learns that $R$ should be constant. Noting that $f(R_{\mu\nu}, g_{\mu\nu})$ is a scalar function in which indices of powers of Ricci tensor $R_{\mu\nu}$ are appropriately summed over by powers of metric $g_{\mu\nu}$, we learn that for the Einstein manifold \emph{vacuum} solutions 
\beq
\frac{\pd f}{\pd R_{\mu \nu}}= \frac{d+1}{R}\frac{\pd f}{\pd g_{\mu \nu}}=\frac{d+1}{2R} f g^{\mu\nu},
\eeq
where the second equality follows from equations of motion,\footnote{This result has been used to study black hole entropy in 3-dim higher curvature gravities \cite{Saida:1999ec}.} and hence

\be
\bg_{\mu\nu}={\cal X}^{\frac{2}{d-1}}\ g_{\mu\nu},\qquad {{\cal X}=\frac{(d+1)f}{2R}}.
\ee
Note that $f$ in the above is the Lagrangian evaluated on \eqref{Einstein-manifold} and is a constant. Comparing the values of on-shell actions one can also relate $\kappa^2$ to $G_{d+1}$.

We may now use the RT prescription to obtain the  functional for $f$(\text{Ric}) theory on Einstein manifold solutions
\beq \label{eRT-f(Rici)}
S_{EE} =\frac{1}{4G_{d+1}} \int_{\gamma_{A}}d^{d-1}\xi \sqrt{\bar\gamma}= \frac{{\cal X}}{4G_{d+1}} \int_{\gamma_{A}}d^{d-1}\xi \sqrt{\gamma}.
\eeq

As a cross-check one can show that this result produces the correct universal term of entanglement entropy. To see this, we recall that the coefficient of the universal part of the entanglement entropy in an even dimensional CFT$_d$ is related to central charges appearing in the trace anomaly of energy-momentum tensor, in 2d \cite{Holzhey:1994we,*Calabrese:2004eu} and in 4d \cite{Solodukhin:2008dh}.\footnote{The story is complicated in 6d CFTs (see \cite{Hung:2011xb , Safdi:2012sn, *Miao:2015iba}).} The form of \eqref{eRT-f(Rici)} functional for $f$(Ric) theory shows this coefficient is the same as Einstein gravity up to an overall constant factor proportional to $f$(Ric)$\mid_{\mathrm{AdS}}$. On the other hand, it is known that $A$-type trace anomaly is proportional to the on-shell value of Lagrangian  evaluated on AdS$_{d+1}$ background, $f$(Ric)$\mid_{\mathrm{AdS}}$ \cite{Myers:2010tj,*Myers:2010xs}. It is also known the type of anomaly which appears in the universal term depends on the shape of entanglement region.\footnote{See also \cite{Bueno:2015rda} for related studies about universal properties of CFTs captured by entanglement entropy.} For example in a CFT$_{4}$, sphere and cylinder return $a$ and $c$ in entanglement entropy, respectively \cite{Solodukhin:2008dh}. However, as we already mentioned, only the $a$ central charge which is related to $A$-type anomaly, can be appear in \eqref{eRT-f(Rici)}. So, the form of functional suggests two central charges $c$ and $a$ are the same for a CFT$_4$ theories which are dual to $f$(Ric) gravity. Therefore, one may conclude only Riemann tensor contributes in $c-a$. This result is consistent with previous studies \cite{Buchel:2008vz, Miao:2013nfa}.

\section{Comparison with other proposals}\label{sec-4}

As discussed in the introduction, there are some different proposals to extend the RT functional for computation of the HEE to higher derivative gravity theories, e.g. see \cite{Fursaev:2013fta,Dong:2013qoa,Camps:2013zua,Miao:2014nxa}. One class of such proposals simply suggest to promote the Wald entropy formula \cite{Wald:1993nt,Iyer:1994ys} to the HEE functional. This proposal is based on the observation that (1) the RT minimal surface has a vanishing trace of extrinsic curvature and that (2) the bifurcation surface of the Killing horizon of a stationary black hole has also a vanishing extrinsic curvature, and that for the latter one can use Wald formula to compute the black hole entropy. If one could show or argue that for a generic higher derivative theory the minimal surface should always have a vanishing extrinsic curvature, then this idea could be a valid one. However, there is a priori no reason that this latter condition should hold. So, in general one would expect the HEE functional should have some extra terms compared to the Wald formula.   In particular it has been shown that this naive idea does not always produce  the correct universal part of entanglement entropy \cite{deBoer:2011wk, Hung:2011xb}.

In the context of black hole entropy in higher derivative gravity theories, there is another functional by Jacobson-Myers \cite{Jacobson:1993vj}. The JM functional which has been proposed for Lovelock gravity theories, has a term proportional to the extrinsic curvature of the codimension two surface over which we are integrating. This term vanishes at the bifurcation surface of the horizon and the JM functional reduces to the Wald entropy formula. It has been shown that using this as an HEE functional yields the correct universal term in the HEE for Lovelock theories \cite{deBoer:2011wk, Hung:2011xb}. 

The  general HEE functional which has been proposed is the one by Xi Dong \cite{Dong:2013qoa} which is an extension of the JM entropy functional.\footnote{See \cite{Miao:2014nxa} for a more general functional.} This functional is given by
\begin{align}\label{Dong-functional}
\begin{split}
S_{\mathrm{Dong}}= 2\pi\int d^{d-1}\xi \sqrt{\gamma} \lt\{ -\fr{\pa L}{\pa R_{\m\r\n\s}} \ve_{\m\r} \ve_{\n\s} + \sum_\a \(\fr{\pa^2 L}{\pa R_{\m_1\r_1\n_1\s_1} \pa R_{\m_2\r_2\n_2\s_2}}\)_\a \fr{2K_{\l_1\r_1\s_1} K_{\l_2\r_2\s_2}}{q_\a+1} \times \rt. \\
\lt.\phantom{\frac12} \[ (n_{\m_1\m_2} n_{\n_1\n_2}-\ve_{\m_1\m_2} \ve_{\n_1\n_2}) n^{\l_1\l_2} + (n_{\m_1\m_2} \ve_{\n_1\n_2}+\ve_{\m_1\m_2} n_{\n_1\n_2}) \ve^{\l_1\l_2}\] \rt\} \,.
\end{split}
\end{align}
where
\begin{align}
\begin{split}
n_{\m\n} = n^{i}_\m n^{j}_\n G_{ij}&,\ \qquad
\ve_{\m\n} = n^{i}_\m n^{j}_\n \ve_{ij}\\
\ve_{\m\n} \ve_{\r\s} = n_{\m\r} n_{\n\s} - n_{\m\s} n_{\n\r} &,\qquad 
K_{\l\m\n} = n^{i}_\l m^{a}_\m m^{b}_\n \mathcal{K}_{ab}^i \,,
\end{split}
\end{align}
with $G_{ij}$ is the metric defined on the two directions perpendicular to the minimal surface, $\ve_{ij}$ is the Levi-Civita tensor, $m^{a}_\m$ are orthogonal unit vectors along the minimal surface directions, $\mathcal{K}^i_{ab}$ is the extrinsic curvature of the minimal surface and also $\a$ runs over different terms in the theory and $q_\a$ is a numerical factor depending on each term which is irrelevant to our analysis. The first term in \eqref{Dong-functional} is the Wald entropy formula, which as discussed, does not depend on the extrinsic curvatures of the minimal surface. 
This proposal reduces to the FPS proposal \cite{Fursaev:2013fta} when the Lagrangian has only curvature squared terms. Here we present the FPS functional explicitly. Consider the following theory curvature squared theory of gravity
\beq \label{Action}
I =-\frac{1}{16 \pi G_{d+1}}\int d^{d+1} x\sqrt{g}\left[R+\frac{d(d-1)}{L^{2}}+\frac{L^{2}}{2}(\alpha_{1}R^{2}+\alpha_{2}R_{\alpha\beta}R^{\alpha\beta}+\alpha_3 R_{\mu\nu\alpha\beta} R^{\mu\nu\alpha\beta})\right]\,,\eeq
The holographic entanglement entropy functional for this theory is given by \cite{Fursaev:2013fta}
\beq \label{FPS}
S_{\mathrm{FPS}}=\frac{1}{4 G_{d+1}}\int d^{d-1}\xi \sqrt{\gamma}\left[1+\frac{L^{2}}{{2}}\left(2\alpha_{1}R+\alpha_{2}( R_{\mu\nu}n^{\nu}_{i}n^{\mu}_{i}-\frac{1}{2} \K^{i}\K_{i})+2\alpha_3(R_{\mu\nu\alpha\beta}n^{\nu}_{j}n^{\mu}_{i}n^{\alpha}_j n^{\beta}_i- \K_{\mu\nu}^{i}\K^{\mu\nu}_{i})\right)\right].
\eeq
where $i$ shows two transverse directions to codimension-2 surface and $\K^i$'s denote trace of extrinsic curvature along this two directions. For the static cases we are considering here, one of these two directions is the time direction. In the appendix A we have presented a more detailed comparison between FPS and our functional.

To produce the HEE these functionals are to be computed on an ``appropriate'' extremal surface. There are two methods to find this extremal surface \cite{Dong:2013qoa}: using the ``conical boundary conditions'' method \cite{Lewkowycz:2013nqa}, or minimizing (extremizing)\footnote{As a side remark, note that for $\alpha_2,\alpha_3>0$, the terms proportional to extrinsic curvature in \eqref{FPS} contribute with a negative sign.} the functional, supplemented by ``causality constraints'' \cite{Headrick:2014eia, Erdmenger:2014tba}.\footnote{For another related prescription see \cite{Hosseini:2015vya}.} Although, it is believed that these two methods should lead to the same extremal surface, to our knowledge there is no generic and ambiguity-free proof of it \cite{Chen:2013rcq,Bhattacharyya:2014yga}. Moreover, it has been shown that the HEE computed with the RT proposal satisfies subadditivity constraints \cite{Headrick:2007km}. It is straightforward to generalize arguments of \cite{Headrick:2007km} and show that this result also holds for {any} HEE functional with extremization method. We, however, note that only one of the extremizing surfaces leads to the correct entanglement entropy.

One may work out the equations governing the extremal surface from variation of  the functional \eqref{Dong-functional}. The details of such a computation may be found in \cite{Bhattacharyya:2014yga} and we have reviewed in the appendix A. 
This equation is generically an order four equation, involving  Laplacian and higher powers of  the extrinsic curvature. Here we give the result for the FPS functional \eqref{FPS} in the $\alpha_3=0$ case, for the Einstein manifolds \eqref{Einstein-manifold}:
\begin{align}\label{eq:eomFPS}
\begin{split}
\left[\frac{2}{L^2}+2\left(\a_1+\frac{\a_2}{d+1}\right)R-\frac{\a_2}{2} \mathcal{K}^{j}\mathcal{K}_{j}+\a_2\nabla_{a}\nabla^{a}\right]\mathcal{K}_{i} +\a_2\left(\mathcal{K}^{ab}_j\mathcal{K}_{iab} +R_{\mu\nu\rho\sigma}\gamma^{ab}n^{\mu}_{i}e^{\nu}_{a}n^{\rho}_{j}e^{\sigma}_{b}\right)\mathcal{K}^{j}=0.
\end{split}
\end{align}
As we see $\mathcal{K}^{j}=0$ is a consistent solution, while there are other solutions too. One may argue that these other solutions do not produce the correct value for entanglement entropy (the anomaly coefficients).

One may repeat  similar analysis for the general case. Although a complicated equation, one may check that for the Einstein manifold solutions to the gravity theories which do not involve Riemann curvature, i.e. the  $f(R,R_{\mu\nu})$ family,  surfaces with vanishing  trace of extrinsic curvature are always a solution (see the appendix A.2).  In our method which is based on mapping to Einstein theory, this latter is a direct outcome.  We have indeed checked explicitly these statements for many different examples in various different theories. We have presented some of those examples in the Appendix B.

\section{Summary and discussion}\label{sec-5}

In this work we analyzed the implications of field redefinition invariance of physical observables in the local quantum field theories and their possible gravity duals, for the entanglement entropy. We used this idea as a method fix the proposal for computing entanglement entropy for field theories with gravity duals. Such gravity duals may in general be a higher derivative gravity theory.  Using the fact that any theory of gravity described by $f(R,R_{\mu\nu})$ Lagrangian can be (formally) mapped onto an Einstein-Hilbert theory upon a field redefinition, and the RT proposal for the computation of entanglement entropy in Einstein-Hilbert gravity, we found the recipe for computing entanglement entropy holographically for $f(R,R_{\mu\nu})$ theories. 

We discussed that our ``field redefinition based'' proposal passes the two tests any such HEE functional is expected to pass:\\
i) it should reduce to the Wald formula when computed on horizon bifurcation surfaces,\\
ii) it should correctly reproduce the universal terms expected to be present in the (H)EE expression.\\
As a check for our analysis, we discussed that our functional reproduces the results of other existing proposals e.g.\cite{Dong:2013qoa, Fursaev:2013fta, Camps:2013zua}, for the class of $f(R,R_{\mu\nu})$ theories, and hence passes these two tests.

Our proposal, among other things, implies that for the class of $f(R,R_{\mu\nu})$ theories the surface which minimizes the corresponding functional (minimal surface) should always have vanishing trace of extrinsic curvature and that for these cases the Dong functional should  reduce to the Wald entropy formula.\footnote{We could only explicitly prove this latter for the existing HEE proposals on Einstein manifold backgrounds. However, our method implies that this result should be true for more general backgrounds solutions of $f(R,R_{\mu\nu})$ theories.} Our arguments would hence provide a new insight on the analysis of 
\cite{Erdmenger:2014tba} and the intricate issues with the choice of minimal surfaces in the higher derivative theories.

Our method, as discussed here, does not in general cover the most general fourth order gravity theories, and in particular the interesting case of Lovelock theories, as the latter always involves powers of Riemann as well. However, we would like to make two comments on such cases: 
\begin{itemize}
\item  Since Riemann curvature tensor can be written in terms of Ricci tensor in three dimensional cases, our method captures the most general parity-conserving three dimensional case.\footnote{In three dimensions we can have the gravitational Chern-Simons term, e.g. as in chiral gravity theories \cite{Deser:1982vy,*Deser:1981wh}. Our method which maps theories to Einstein-Hilbert, will only capture the parity-even parts, like the NMG theory \cite{Bergshoeff:2009hq} which is discussed in the appendix.}  However, we expect it should be possible to extend our method to the general cases with both parity even and odd parts, where the even-parity part of any 3d gravity theory is mapped onto the Einstein-Hilbert and the odd-parity part to simple gravitational Chern-Simons. We shall explore this in future works.
\item It is known that $f(R,R_{\mu\nu})$ theories in the linearized regime, generically, involve a massless  and a massive spin two degrees of freedom. Our field redefinition, was used to disentangle these two degrees of freedom such that the massless graviton is essentially described by our new metric in the Einstein frame.\footnote{It is also known that such higher derivative theories of gravity generically have ghosts. With our mapping, we are choosing the overall sign of our action such that the massless graviton is not the ghost degree of freedom. In our analysis this is reflected in the fact that  in the Einstein frame the Newton constant has the positive sign.} Presence of a massless and a massive spin two degree of freedom is also a generic property of more general $L(R,R_{\mu\nu},R_{\nu\mu\alpha\beta})$ theories.\footnote{ Lovelock theories are special in the sense that, by construction, the field equations are second order and hence they only involve a massless graviton in the linearized level.} One would hence in principle expect that the field redefinition method should also work for more general theories. It would be desirable to explore this direction.    
\end{itemize}

One of the implicit features of the Wald entropy formula or the HEE functional proposals, like the RT or that of Dong, 
is that the entropy only involves the gravitational part of the action. Moreover, our analysis reveals another interesting fact: for the Einstein-like theories, this is only the ``massless spin-two'', the Einstein-Hilbert part, which determines the entropy. On the other hand, the ``matter'' part would appear through the 
background metric.  However, generically, one may have theories in which there is no clear distinction between the gravity and the matter parts. A simple example is theories involving the conformal mass term. Another class which has recently attracted some interest is the Horndeski theories \cite{Horndeski:1974wa}, where the field equations are still second order and the theory is ghost-free. It is interesting to explore the HEE proposals, and in particular, our field redefinition method, for these theories. 

Finally, we would like to point out that, although in our analysis here we only considered entanglement entropy, other information theoretic measures, like relative entropy, mutual information, multi-partite entanglement, should also be 
field redefinition invariant. It would be interesting to explore implications of field redefinition invariance for these other observables.

\subsection*{Acknowledgement}

We would like to thank  Aninda Sinha for reading the draft and useful comments.
The work of M.M. Sh-J is supported in part by Allameh Tabatabaii Prize Grant of Boniad Melli Nokhbegan of Iran, the SarAmadan grant of Iranian vice presidency in science and technology and the ICTP network project NET-68. M.M.Sh-J. is also acknowledging the ICTP Simons fellowship. The authors thank the hospitality of ICTP, Trieste, while this project completed. M.R.M.M., A.M. and M.H.V are supported by Iran National Science Foundation (INSF).

\appendix

\section{More detailed comparison with other proposals}

In this appendix we provide more details of computations needed for the comparison of our results and the FPS and Dong proposals. We will first present the simpler case of FPS proposal which is suited for curvature-squared gravity theories and then discuss the more general case of Dong proposal. As mentioned, the Dong proposal reproduces the FPS functional. Our goal in this appendix is to show that surfaces with $\mathcal{K}=0$ are always solution to the equations of motion for extremal surfaces derived from the FPS or Dong functionals for Einstein manifold background solutions to the $f(R,R_{\mu\nu})$ gravity theory.

\subsection{Comparison with the FPS proposal}

As the starter, let us consider the $\alpha_2=\alpha_3=0$ case, where we deal with an $f(R)$ theory with  $f(R)=R+\frac{d(d-1)}{L^{2}}+\frac{L^{2}\alpha_{1}}{2} R^2$. One may readily show that the FPS functional \eqref{FPS} is exactly equal to \eqref{func-fofr}. 

For the more general case of $\alpha_2\neq 0$ (still $\alpha_3=0$), one can again directly check this by minimizing the functional \eqref{func-fofr} 
 and apply the methods of section \ref{sec-3.2}. Below, we present a sample such computation.


\paragraph{Perturbative analysis.} Recall that in this case we had
$$
\tilde{\cal F}_{\mu\nu}=L^2(\alpha_1 R g_{\mu\nu}+ \alpha_2 R_{\mu\nu})
$$
and hence,
\be
\tilde{\cal F}=L^2(\alpha_2+(d+1)\alpha_1)R,\qquad \gamma^{ab}\tilde{\cal F}_{ab}=L^2[(d-1)\alpha_1 R+ \alpha_2\gamma^{\mu\nu}R_{\mu\nu}],
\ee
where $\gamma^{\mu\nu}=g^{\mu\nu}-n_i^\mu n_i^\nu$. Therefore, \eqref{pertFunctional} becomes
\begin{eqnarray} \label{pertFuncR2}
S_{EE}=\frac{1}{4G_{d+1}} \int_{{\gamma_{A}}}d^{d-1}\xi\sqrt{\gamma}\left[1+ { L^2}\left(\alpha_{1}R+\frac{\alpha_{2}}{2}R_{\mu\nu}n^{\nu}_{i}n^{\mu}_{i} \right)\right]+\mathcal{O}(\alpha_i^2).
\end{eqnarray}
In the absence of the anomalous term (trace of extrinsic curvature), the above result is the same as FPS functional \eqref{FPS}. One can show that the contribution of the extrinsic curvature terms will always be of the second order in $\alpha_i$. The argument is as follows: 
As we already pointed out the trace of extrinsic curvature of the minimal surface is zero for the Einstein theory and in the absence of higher derivative corrections \cite{Hubeny:2007xt} and hence $\K_i$ will necessarily be (at least) of the order $\alpha_i$. Therefore, the last term in \eqref{FPS} will not contribute to the leading order analysis.\footnote{For higher curvature theory in general the extremal surface will not necessarily have vanishing extrinsic curvature $\K_i$. Note also that as was shown in \cite{Chen:2013rcq, Bhattacharyya:2014yga} the surface extremizing the FPS functional does not necessarily coincide with
the surface determined by the generalized entropy method of \cite{Lewkowycz:2013nqa}. Some authors choose a specific limiting procedure to match these two methods at least for Lovelock theories \cite{Bhattacharyya:2013gra,Dong:2013qoa}. See \cite{Camps:2014voa} for another approach to this problem.}

\paragraph{Einstein manifold backgrounds.}
Although one may study generic backgrounds under field redefinition invariance, for technical simplicity we restrict our analysis to Einstein manifold backgrounds. These backgrounds include all interesting well-known geometries previously studied in the literature such as pure AdS and AdS-Schwarzschild. For such backgrounds by choosing $f(\text{Ric})=R+\frac{d(d-1)}{L^{2}}+\frac{L^{2}}{2}(\alpha_{1}R^{2}+\alpha_{2}R_{\alpha\beta}R^{\alpha\beta})$ the FPS functional becomes
\be
S_{\mathrm{FPS}}=\frac{1}{4 G_{d+1}}\int d^{d-1}x \sqrt{\gamma}\left[1+\frac{L^{2}}{{2}}\left(2\alpha_{1}R+\alpha_{2}\left(\frac{R}{d+1} g_{\mu\nu}n^{\nu}_{i}n^{\mu}_{i}-\frac{1}{2} \K^{i}\K_{i}\right)\right)\right],
\ee
where for any time-like or space-like hyper-surface $g_{\mu\nu}n^{\nu}_{i}n^{\mu}_{i}=2$. On the other hand using equations \eqref{fEq} and \eqref{eRT1} one may trivially arrive at
\beq \label{eRTEM}
S_{EE} =\frac{1}{4G_{d+1}} \int_{\gamma_{A}}d^{d-1}\xi \sqrt{\gamma}\left[1+L^{2}\left(2\alpha_{1}R+\alpha_{2}\frac{R}{d+1}\right)\right].
\eeq

\paragraph{Minimal surface for FPS.}
Here we study the equation of the extremal surface to show that it always admits $\mathcal{K}^{i}=0$
as a solution. Such an equation is found by extremizing the FPS functional (\ref{FPS}). Using the results of section 3 of \cite{Bhattacharyya:2014yga}, all the ingredients needed to calculate variation of (\ref{FPS}) are
\begin{align}\label{eq:varFPS}
\begin{split}
\delta\sqrt{\gamma}&=\sqrt{\gamma}\mathcal{K}_i\psi^i,\\ 
\delta\left(\sqrt{\gamma}R\right)&=\sqrt{\gamma}\left(\mathcal{K}_iR+n_i^\mu\hat{\nabla}_\mu R\right)\psi^i, \\
\delta\left(\sqrt{\gamma}R_{\mu\nu}n^{\nu j}n_j^\mu\right)&=\sqrt{\gamma}\left(\mathcal{K}_iR_{\mu\nu}n^{\nu j}n_j^\mu+2\nabla^a\left(R_{\mu\nu}n_i^\nu e_a^\mu\right)-n_i^\sigma \gamma^{ab}e_a^\mu e_b^\nu \hat{\nabla}_\sigma R_{\mu\nu}+n_i^\mu \hat{\nabla}_\mu R \right)\psi^i,\\
\delta(\sqrt{\gamma} \mathcal{K}^{i}\mathcal{K}_{i})&=-2\sqrt{\gamma}\left(\nabla_{a}\nabla^{a}\mathcal{K}_{i}-\frac{1}{2} \mathcal{K}^{j}\mathcal{K}_{j}\mathcal{K}_{i} +\mathcal{K}_{j}\mathcal{K}^{jab}\mathcal{K}_{iab} +\mathcal{K}^{j}R_{\mu\nu\rho\sigma}\gamma^{ab}n^{\mu}_{i}e^{\nu}_{a}n^{\rho}_{j}e^{\sigma}_{b} \right)\psi^{i},
\end{split}
\end{align}
where following the conventions of \cite{Bhattacharyya:2014yga} $\hat{\nabla}$ denotes the covariant derivative contemptible with bulk metric and $\psi^{i}$ is the variation parameter.

For the case of Einstein manifolds (\ref{Einstein-manifold}), $\hat{\nabla}_{\rho}R_{\mu \nu}=0$ and the two middle equations reduce to
\begin{eqnarray}
\delta\left(\sqrt{\gamma}R\right)&=&\sqrt{\gamma}\mathcal{K}_iR \psi^i, \\
\delta\left(\sqrt{\gamma}R_{\mu\nu}n^{\nu j}n_j^\mu\right)&=&\sqrt{\gamma}\mathcal{K}_iR_{\mu\nu}n^{\nu j}n_j^\mu\psi^i.
\end{eqnarray}
Using these ingredients with the appropriate coefficients leads to the minimal surface equation \eqref{eq:eomFPS}.
It is easy to see that \eqref{eq:eomFPS} always admits a $\mathcal{K}^{i}=0$ solution. Thus for Einstein manifolds the functional (\ref{eRTEM}) exactly matches with (\ref{FPS}). On the other hand for the case of general manifolds we have already mentioned that due to our perturbative analysis, the extrinsic curvature contributions are at least of the order $\alpha_{i}$ and thus will not contribute to leading order of entropy. 

Some comments are in order:
\begin{itemize}

\item The equation of surface \eqref{eq:eomFPS} may admit another solution which we have not considered in our analysis. However, the $\mathcal{K}^{i}=0$ solution leads to the correct universal term for EE and so we are considering it as the appropriate choice.

\item Lewkowycz and Maldacena \cite{Lewkowycz:2013nqa} have used a different approach to obtain the equation of the minimal surface. They consider field equations of the bulk metric in the vicinity of a small conical deficit and by demanding the matter stress-energy tensor to be finite, they set the coefficients of divergent terms to vanish. This condition fixes the location of the extremal surface. As it is argued in \cite{Dong:2013qoa} we expect this procedure to be the same as variation of the EE functional.\footnote{However, as it is mentioned in \cite{Chen:2013rcq,Dong:2013qoa,Bhattacharyya:2014yga,Camps:2014voa} there are some ambiguities in taking the corresponding limits which we are not going to discuss here.} In the case of \eqref{FPS} one may show that $\mathcal{K}^{i}=0$ is one of the Lewkowycz-Maldacena conditions \cite{Bhattacharyya:2014yga}.
\end{itemize}

\subsection{Comparison with the Dong proposal}
We are specifically interested to compute the second `anomalous' terms in the case of $f(\mathrm{Ric})$ theories. To do so use the following identities
\begin{align}
\begin{split}
\fr{\pa^2 f(\mathrm{Ric})}{\pa R_{\m_1\r_1\n_1\s_1} \pa R_{\m_2\r_2\n_2\s_2}}&=
\fr{\pa R_{\mu\nu}}{\pa R_{\m_1\r_1\n_1\s_1}}\fr{\pa R_{\rho\sigma}}{\pa R_{\m_2\r_2\n_2\s_2}}
\fr{\pa^2 f(\mathrm{Ric})}{\pa R_{\m\n}\pa R_{\r\s}},\\
\fr{\pa R_{\alpha\beta}}{\pa R_{\m\r\n\s}}&=
\tfrac{1}{8} (-\delta_{\alpha }{}^{\nu } \delta_{\beta }{}^{\rho } \
\mathit{g}^{\mu \sigma } + 2\delta_{\alpha }{}^{\rho } \delta_{\beta \
}{}^{[\sigma} \mathit{g}^{\nu]\m}- \delta_{\alpha }{}^{\mu } \delta_{\beta \
}{}^{\sigma } \mathit{g}^{\nu \rho } + 2\delta_{\alpha }{}^{\sigma }
\delta_{\beta }{}^{[\rho } \mathit{g}^{\mu]\nu } + 2\delta_{\alpha }{}^{(\nu } \
\delta^{\mu)}_{\beta }{}\mathit{g}^{\rho \sigma }).
\end{split}
\end{align}

We again restrict our analysis to the special case of Einstein manifolds. In such a case due to the symmetry properties w.r.t. $\{\m, \n\}$ and also  $\{\r, \s\}$ and w.r.t. these two sets we introduce the following notations
\begin{align}
\begin{split}
\fr{\pa^2 f(\mathrm{Ric})}{\pa R_{\m\n}\pa R_{\r\s}}&=\mathcal{A}^{\mu\nu}\mathcal{B}^{\r\s}\\
\mathcal{A}_{ij}&=\mathcal{A}_{\mu\nu}n^\m_{i}n^\n_{j}
\end{split}
\end{align}
which leads to the following result for the anomalous terms
\begin{align}
\begin{split}
\fr{\pa^2 f(\mathrm{Ric})}{\pa R_{\m_1\r_1\n_1\s_1} \pa R_{\m_2\r_2\n_2\s_2}} &K_{\l_1\r_1\s_1} K_{\l_2\r_2\s_2}
\[ (n_{\m_1\m_2} n_{\n_1\n_2}-\ve_{\m_1\m_2} \ve_{\n_1\n_2}) n^{\l_1\l_2} + (n_{\m_1\m_2} \ve_{\n_1\n_2}+\ve_{\m_1\m_2} n_{\n_1\n_2}) \ve^{\l_1\l_2}\]\\&=
-\frac{(\mathcal{K}_i\mathcal{K}^i)}{16}\sum_{i,j=1}^2\left(\mathcal{A}_{ii}\mathcal{B}_{jj}-\frac{1}{2}\mathcal{A}_{ij}\mathcal{B}_{ij}\right).
\end{split}
\end{align}
For the case of Einstein manifolds the above summation leads to a $R$-dependent constant value which does not participate in the variation of the entropy functional. Since the Wald contribution is independent of the extrinsic curvature of the minimal surface, the above contribution to the minimal surface equation as $\delta(\sqrt{\gamma} \mathcal{K}^{i}\mathcal{K}_{i})$ which was previously calculated in \eqref{eq:varFPS}. 
One can hence deduce that $\mathcal{K}_i=0$ is a solution
for the minimal surface equation  for the case of Einstein manifolds in support of our result in \eqref{eRTEM}.

\section{Explicit Examples}

In this appendix we work out the entanglement entropy for specific examples in higher derivative theories for different entangling regions with our field redefinition method. We compare our results with the existing results in the literature. This will show how our method works in practice and also provides direct and explicit checks for our proposal. Here we only consider examples of $f$(Ric) theories.

\subsection{$f(\text{Ric}) =R-2\Lambda +\alpha_{1} R^2+\alpha_{2} R_{\mu \nu}R^{\mu \nu}$}\label{sec-4.2}
Considering the following AAdS solution 
\beq\label{eq:AAdSdp1}
ds^2=\frac{\tilde{L}^2}{z^2}\left(b(r)d\tau^2+\frac{dz^2}{b(r)}+h_{ij}dx^{i}dx^{j} \right),
\eeq
where $b(z)=1-m\,z^{d-1}$, $\tilde L=\frac{L}{\sqrt{f_{\infty}}}$ denotes the radius of AdS space, $h_{ij}$ is a $(d-1)$-dimensional Euclidean metric and $f_\infty$ satisfies the following equation
\beq\label{LLtilde}
1-f_{\infty}+\frac{d(d-3)}{2(d-1)}\left[(d+1)\alpha_1+\alpha_2\right]f_{\infty}^2=0.
\eeq

In the case of Ricci-squared theory \eqref{Action} and when we encounter asymptotically AdS$_{d+1}$ solutions, one can simply show that functional \eqref{eRT-f(Rici)} is given by
\beq\label{eRT}
S_{EE}=\frac {{\cal X}}{4G_{d+1}} \int_{\gamma}d^{d-1}\xi \sqrt{\gamma},
\eeq
where
\begin{eqnarray}
{\cal X}\equiv\frac {(d+1)f(\text{Ric})|_{AAdS}}{2R}
=1-d(\alpha_{1}(d+1)+\alpha_{2})f_{\infty},
\end{eqnarray}
where we have used \eqref{LLtilde} to obtain the second equality.
Naturally, one may show \eqref{eRT} is the same as perturbative functional \eqref{pertFuncR2} after a perturbative expansion. In the following we will use this functional to compare our result with FPS functional \eqref{FPS}.

\paragraph{Spherical entangling region.}
In order to study a spherical entangling region we consider the $h_{ij}$ part of the background \eqref{eq:AAdSdp1} as
\beq
h_{ij}dx^idx^j=d\rho^2+\rho^2d\Omega_{d-2}^2.
\eeq
We consider a spherical entangling, i.e., $\rho=\ell$ region at constant time slice on the boundary. In what follows we treat the black hole effects perturbatively at leading order of $m$ thus we parametrize the RT surface in the bulk as $\rho(z)=g_0(z)+m\,g_1(r)$. For this configuration the FPS functional \eqref{FPS} reads as
\begin{align}
\begin{split}
S_{\mathrm{FPS}}=\frac{(d-1)\mathrm{Vol}(S^{d-1})\tilde{L}^{d-1}}{4G_N}&\int dz\; \frac{g_0^{d-2} \sqrt{1+{g_0'}^2}}{z^{d-1}}\times
\\&\hspace{-13mm}\left[{\cal X}-\frac{\alpha_2f_{\infty} \left((1+{g_0'}^2) \left((d-1) g_0 g_0'+(d-2) z\right)-zg_0 g_0''\right)^2}{4 g_0^2 (1+{g_0'}^2)^3}+m\mathcal{F}_d\left[g_0,g_1\right]+\mathcal{O}\left(m^2\right)\right],
\end{split}
\end{align}
where we are not interested in the explicit (messy) form of $\mathcal{F}_d\left[g_0,g_1\right]$ before solving for the profile of $\rho(z)$. The equation of motion for $g_0(z)$ and $g_1(z)$ could be solved by
\be\label{profSph}
g_0(z)=\sqrt{\ell^2-z^2}\;\;\;\;\;,\;\;\;\;\;g_1(z)=\frac{2\ell^{d+2}-\ell^2r^d-r^{d+2}}{2(d+1)\sqrt{\ell^2-z^2}}.
\ee
Substituting back the explicit form of this profile into the FPS functional one finds
\begin{align}\label{FPSddim}
\begin{split}
S_{\mathrm{FPS}}=
\frac{(d-1)\mathrm{Vol}(S^{d-1})\tilde{L}^{d-1}}{4G_N}\,{\cal X}
&\int dz\frac{(\ell^2-z^2)^{\frac{d-3}{2}}}{z^{d-1}}\times\\&\hspace{-8mm}\left[1+\left(2(d-2)\ell^{d+2}-2\ell^dr^2+(d+3)\ell^2r^d+3(d-1)r^{d+2}\right)m+\mathcal{O}\left(m^2\right)\right].
\end{split}
\end{align}
On the other hand our functional \eqref{eRT} for the same configuration reads
\begin{align}\label{eRTSph}
\begin{split}
S_{EE}=\frac{(d-1)\mathrm{Vol}(S^{d-1})\tilde{L}^{d-1}}{4G_N}{\cal X}\int dz\;& \frac{g_0^{d-2} \sqrt{1+{g_0'}^2}}{z^{d-1}}\times\\&\left[1+\frac{2(d-2)g_1(1+{g'_0}^2)+g_0(z^d+2g'_0g'_1)} {2g_0(1+{g'_0}^2)}m+\mathcal{O}\left(m^2\right)\right].
\end{split}
\end{align}
It is not hard to show that the same profile \eqref{profSph} minimizes \eqref{eRT} thus implementing it into \eqref{eRTSph} leads to
the same result which we found from $S_{\mathrm{FPS}}$.
\paragraph{Infinite strip entangling region.}
In order to study an infinite strip entangling region we consider the $h_{ij}$ part of the background \eqref{eq:AAdSdp1} as
\beq
h_{ij}dx^idx^j=\sum_{i=1}^{d-1}dx_i^2.
\eeq
The entangling region is defined as
$$\tau=\mathrm{const.}\;\;\;\;,\;\;\;\; -\frac{\ell}{2}<x_1<\frac{\ell}{2}\;\;\;\;,\;\;\;\;-L_a<x_a<L_a,\;\;\mathrm{for}\;\;a=2,\cdots, d-1,$$
and since we again treat the black-hole effects perturbatively at leading order of $m$, we parametrize the RT surface in the bulk as $x_1\equiv g(z)=g_0(z)+m\,g_1(r)$. The FPS functional \eqref{FPS} becomes
\begin{align}\label{fpsstrip}
S_{\mathrm{FPS}}=&\frac{V_{d-1}\tilde{L}^{d-1}}{4G_N}\int dz\; \frac{\sqrt{1+{g'}^2}}{z^{d-1}}\left[{\cal X}-\frac{\alpha_2f_{\infty} \left((1-d)g'(1+{g'}^2) +z g''\right)^2}{4 (1+{g'}^2)^3}+m\mathcal{F}_d\left[g_0,g_1\right]+\mathcal{O}\left(m^2\right)\right].
\end{align}
It is not hard to find the following profile for the minimal surface
\begin{align}\label{profilestrip}
\begin{split}
g_0(z)&=\frac{z_t}{d}\left(\frac{z}{z_t}\right)^d{}_2F_1\left[\frac{1}{2},\frac{d}{2(d-1)},1+\frac{d}{2(d-1)},\left(\frac{z}{z_t}\right)^{2(d-1)}\right]\\
g_1(z)&=\frac{z^{2d}-z^2z_t^{2(d-1)}}{2(d+1)\sqrt{z_t^{2(d-1)}-z^{2(d-1)}}}+\frac{z^2z_t^{d-1}}{2(d+1)}{}_2F_1\left[\frac{1}{d-1},\frac{1}{2},1+\frac{d}{d-1},\left(\frac{z}{z_t}\right)^{2(d-1)}\right]
\end{split}
\end{align}
where $z_t$ is the turning point and it is fixed by the boundary data $\ell$ by
$$z_t=\frac{\Gamma(\frac{1}{2d-2})}{\sqrt{\pi}\Gamma(\frac{d}{2d-2})}\frac{\ell}{2}.$$
Plugging this profile into FPS functional shows that similar to what happens for the spherical entangling region, the extrinsic curvature term does not contribute to FPS functional thus we have
\begin{align}\label{fpsstrip2}
S_{EE}=S_{\mathrm{FPS}}=&\frac{V_{d-1}\tilde{L}^{d-1}}{4G_N}{\cal X}\int dz\; \frac{z_t^{d-1}}{z^{d-1}}\frac{1}{\sqrt{z_t^{2d-2}-z^{2d-2}}}\left[1+\frac{1}{2}z^d\,m+\mathcal{O}\left(m^2\right)\right].
\end{align}

\paragraph{Infinite cylinder entangling region.}
As our last case we study an infinite cylinder entangling region. We consider the $h_{ij}$ part of the background \eqref{eq:AAdSdp1} as
\beq
h_{ij}dx^idx^j=du^2+d\rho^2+\rho^2d\Omega_{d-3}^2.
\eeq
The entangling region is defined as
$$\tau=\mathrm{const.}\;\;\;\;,\;\;\;\; \rho=\ell\;\;\;\;,\;\;\;\;0<u<H,$$
where $H$ is an IR regulator.
In this case since we are not able to find the profile analytically, we restrict our analysis to the near boundary region of the pure AdS$_{d+1}$ space and thus we try to find the RT surface assuming $\rho=g(z)$. For such an entangling region the FPS functional \eqref{FPS} is
\begin{align}\label{fpscylinder}
\begin{split}
S_{\mathrm{FPS}}=&\frac{(d-2)\mathrm{Vol}(S^{d-2})\tilde{L}^{d-1}H}{4G_N}\times\\
&\int dz\; \frac{g^{d-3} \sqrt{1+{g'}^2}}{z^{d-1}}\left[{\cal X}-\frac{\alpha_2f_{\infty} \left((1+{g'}^2) \left((d-1) g g'+(d-3) z\right)-zg g''\right)^2}{4 g^2 (1+{g'}^2)^3}\right].
\end{split}
\end{align}
The equation of motion for $g(z)$ can be solved near the boundary $z\rightarrow 0$ as follows
\begin{align}\label{asymprofile}
g(z)=\ell-\frac{d-3}{2(d-1)}\frac{z^2}{\ell}+\mathcal{O}\left(z^4\right).
\end{align}
Indeed, as we have just seen in the case of spherical and infinite strip entangling regions, in this case the extrinsic curvature again not contribute to the on-shell FPS functional and we can compare the result with our functional
\begin{align}\label{ertcylinder}
S_{EE}=&\frac{(d-2)\mathrm{Vol}(S^{d-2})\tilde{L}^{d-1}H}{4G_N}{\cal X}\int dz\; \frac{g^{d-3} \sqrt{1+{g'}^2}}{z^{d-1}},
\end{align}
where the same profile minimizes this functional in the near boundary region. Thus we are lead to
\beq
S_{EE}=S_{\mathrm{FPS}}=\frac{(d-2)\mathrm{Vol}(S^{d-2})\tilde{L}^{d-1}H}{4G_N}{\cal X}\int_{z\sim \epsilon} dz\; \mathcal{S}_d(z),
\eeq
where
\begin{align}
\mathcal{S}_d(z)=
\begin{cases}
\frac{\ell}{z^3}-\frac{1}{8\ell z}&d=4\\
\frac{\ell^2}{z^4}-\frac{4}{9 z^2}-\frac{5}{81\ell^2}&d=5\\
\frac{\ell^3}{z^5}-\frac{27\ell}{z^3}+\frac{135}{2048\ell z}&d=6
\end{cases}.
\end{align}
\subsection{New Massive Gravity}\label{sec-4.3}
Here we consider the NMG action as follows (in order to compare the final results we will use same convention as \cite{Erdmenger:2014tba})
\begin{align}
I_{\rm NMG}=\frac{1}{16\pi G_N}\int d^3x\sqrt{-g}\left(\sigma R-2\lambda m^2+\frac{1}{m^2}\left(R_{\mu\nu}R^{\mu\nu}-\frac{3}{8}R^2\right)\right),
\end{align}
where $\sigma=\pm 1$. The above action has a BTZ solution which is given by
\begin{align}
ds^2=-\left(-M+\frac{r^2}{\ell^2}\right)dt^2+\frac{dr^2}{-M+\frac{r^2}{\ell^2}}+r^2 d\phi^2,
\end{align}
where $\frac{1}{\ell^2}=2m^2\left(\sigma\pm\sqrt{1+\lambda}\right)$. Considering the entangling region in a constant time slice with $-\frac{\phi_0}{2}<\phi<\frac{\phi_0}{2}$ and denoting the profile of the minimal surface\footnote{For previous studies of minimal surfaces in BTZ solutions of NMG see \cite{Erdmenger:2014tba, Ghodsi:2015gna}.} by $\phi=g(r)$ the HEE functional \label{eRT1} becomes 
\be
S_{EE}=\frac{1}{4G_N{\cal X}}\int dr \sqrt{r^2g^{'2}+\frac{1}{-M+\frac{r^2}{\ell^2}}},
\ee
where ${\cal X}=-\frac{4}{\ell^2}\left(\sigma+\frac{1}{2m^2\ell^2}\right)$. The equation of motion for the minimal surface using a conserved quantity becomes
\begin{align}
g'(r)=\pm\frac{r_t\ell}{r\sqrt{(r^2-r_t^2)(r^2-M\ell^2)}},\;\;\; g'(r_t)=\infty.
\end{align}
Finally the HEE becomes
\begin{align}
S_{\rm EE}=\frac{c}{3}\log \frac{2r_{\infty}}{r_t^2-M\ell^2},
\end{align}
where $c=\frac{3\ell}{2G_N}\left(\sigma+\frac{1}{2m^2\ell^2}\right)$. Thus the correct central charge is found without the extra causality constraint previously discussed in \cite{Erdmenger:2014tba}.

\bibliographystyle{fullsort}
\addcontentsline{toc}{section}{References} 
\bibliography{FRHEE2}

\end{document}